# High quality GaMnAs films grown with arsenic dimers

R. P. Campion, K. W. Edmonds, L. X. Zhao, K. Y. Wang, C. T. Foxon, B. L. Gallagher and C. R. Staddon

*School of Physics and Astronomy, University of Nottingham, Nottingham NG7 2RD*

***Abstract:*** *We demonstrate that GaMnAs films grown with $As_2$ have excellent structural, electrical and magnetic properties comparable or better than similar films grown with $As_4$. Using $As_2$, a Curie temperature of 112K has been achieved, which is slightly higher than the best reported to date. However, more significantly films showing metallic conduction have been obtained over a much wider range of Mn concentrations from 1.5 to 8% than has previously been reported for films grown with $As_4$. The improved properties of the films grown with $As_2$ are related to the lower concentration of anti-site defects at the low growth temperatures employed.*

## 1. Introduction

Ferromagnetic semiconductors have the potential to form the basis of new types of Spintronic devices in which the spin degree of freedom of the carries is exploited in, for example, polarisation sensitive opto-electronic devices. Such devices will require materials with good electrical and structural quality combined with high Curie temperatures. Much of the effort in this field has focused on the GaMnAs system, which has been used as the basis of several demonstrator spintronic devices [1]. However, the maximum Curie temperature of 109K achieved for this system [2] is well below that predicted by theory [3]. Furthermore metallic conduction has only been observed in GaMnAs over a very narrow range of Mn concentrations. There would therefore still seem to be considerable potential for improving on the quality of the GaMnAs films produced so far.

To achieve ferromagnetism in GaMnAs requires the incorporation of Mn at levels that are far above the equilibrium solubility limit. It is therefore necessary to grow films at low temperature using Molecular Beam Epitaxy (MBE). The low growth temperatures can lead to the incorporation of high levels of defects, in particular As antisite defects ($As_{Ga}$) [4]. To try to minimise this problem films are usually grown with low arsenic to gallium flux ratio. This improves both the properties of the films and enables RHEED oscillations to be observed thus confirming that a 2D-growth mode is obtained under such condition [5]. It is also found that both the ferromagnetic transition temperature and electrical properties can be improved by post growth annealing. An effect that is probably due to reducing defect levels [4].

Previously almost all GaMnAs films have been grown using tetramers ($As_4$), which are obtained by direct evaporation of arsenic. The incorporation kinetics of arsenic dimers [6] ($As_2$) differ substantially from those of tetramers [7]. In particular $As_2$ has a shorter surface lifetime of compared with $As_4$ molecules at any given growth temperature. This leads to improved properties for GaAs based quantum wells [8] and lasers [9] grown at conventional MBE temperatures. The improvement in properties results from a lower concentration of point defects in films grown with $As_2$ compared to those grown with $As_4$ [10]. GaAs films grown at lower temperatures also have significantly better properties when grown with the dimer form of arsenic [11]. It follows that we may expect GaMnAs films grown at low temperature with $As_2$ to have significantly different and possibly better properties than corresponding films grown with $As_4$.

We have investigated GaMnAs films grown with $As_2$ as a function of growth temperature and Mn flux and we have compared the structural, magnetic and electrical transport properties of such films with similar samples grown with $As_4$ elsewhere.

## 2. Experimental Details

GaMnAs films have been grown by MBE using a modified Varian GEN-II system equipped with conventional Al, Ga (2), Mn, Si, Mg effusion cells together with a two-zone As cracker cell and RF source for active nitrogen. The system is equipped with a conventional reflection high-energy electron diffraction (RHEED) system operating at about 12keV and in a separate analysis chamber, Auger electron spectroscopy (AES) with in-situ depth profiling capability using argon ion sputtering. The growth temperature is measured using a thermocouple arrangement, which monitors the sample temperature and is not sensitive to the heater temperature. The relationship between growth temperature and thermocouple temperature has been measured at high temperature using a pyrometer and the resulting curve for each substrate block has been extrapolated to low temperature to give the quoted growth temperatures. However, the absolute accuracy of this measurement cannot be guaranteed at the low temperatures used for the growth of the active part of the structures and are therefore specific to this equipment.

The films discussed in this paper have the following structure: semi-insulating GaAs substrate/high temperature GaAs buffer layer/low temperature GaAs buffer layer/low temperature GaMnAs active layer, with Mn concentrations in the range 0.5% to 8%. The films are grown on (001) oriented semi-insulating GaAs substrates. The substrates are first heated to approximately 400$^o$C in the buffer chamber to remove water vapour, the oxide is then removed at approximately 580$^o$C under an arsenic flux. Next an approximately 100nm thick, high temperature buffer layer is deposited at 580$^o$C. The samples are then cooled to the growth temperature for the active layer, initially under an arsenic flux down to 400$^o$C, and subsequently without an intentional arsenic flux present. A 50nm thick GaAs buffer layer is then grown at the low growth temperature and the films studied with in-situ RHEED to ensure a two-dimensional (2D) growth mode is obtained. The low temperature buffer layer is essential in order to allow the films to reach thermal equilibrium at the lower temperature, since at this low temperature the samples are heated significantly by the radiation from the high temperature Ga cells. Two Ga cells are used to provide the total Ga flux when growing the low temperature GaAs buffer, one producing a flux which is much larger than the other, the combined fluxes determining the overall growth rate of GaAs. GaMnAs is then grown by closing the shutter for the smaller Ga flux and simultaneously opening the Mn shutter. The smaller Ga flux and Mn flux are adjusted using the in-situ ion gauge to have approximately equal intensity, taking into account the different gauge sensitivity for the two elements. This procedure guarantees that we maintain both the sample temperature and the stoichiometry of the films during the growth of GaMnAs. The growth rate for all the above stages is ~0.3 μm/h. After growth the films are cooled to room temperature or annealed in-situ as described below. Two types of structure have been grown, 1μm thick samples to estimate the Mn content and 50nm thick samples for electrical measurements. Low temperature GaAs samples, grown at the same temperature as the GaMnAs films, have also been prepared using the same methods for structural assessment.

The structural properties of the GaMnAs and GaAs films were determined using X-ray diffraction studies in a Philips X'Pert MRD system. For most samples, reciprocal space maps using symmetric 004 and asymmetric 444 reflections have been studied to determine the degree of relaxation and the lattice parameter, relative to the underlying GaAs substrate. Glancing incidence X-ray diffraction measurements have also been used to study thin GaMnAs films.

Magneto-transport measurements have been carried out on Hall bar samples in the temperature range 0.3 to 300K and in magnetic fields up to 16 Tesla. This enables us to separate the spontaneous and normal contributions to the Hall resistance and thus obtain the carrier density and a measure of the magnetisation of the samples.

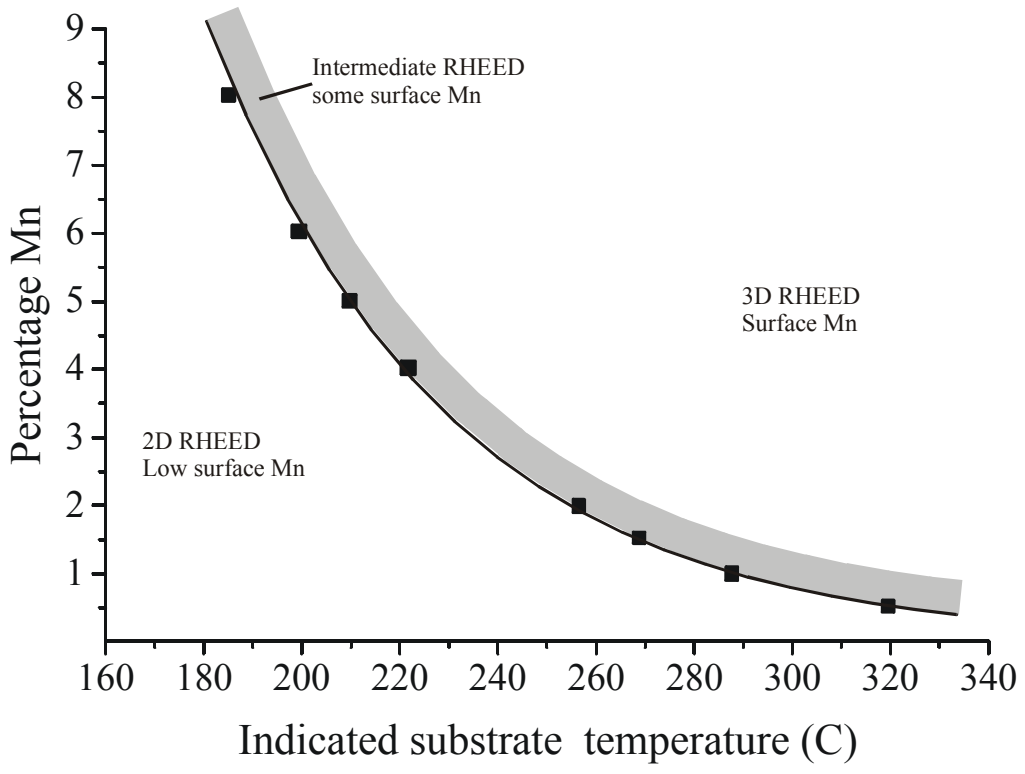

Figure 1 The transition from a 3D to 2D growth mode for GaMnAs films grown with $As_2$ as a function of growth temperature and Mn flux.

## 3. Results

### 3.1. The growth modes

We have studied the growth of GaMnAs films as a function of Mn flux and growth temperature. As previously reported [12], at high Mn fluxes or high growth temperatures, a three-dimensional (3D) growth mode is observed by RHEED. We note that the resulting films appear darker than GaAs, but do not show evidence for MnAs inclusions. At lower Mn fluxes or reduced growth temperature a 2D growth mode is observed and the films appear normal. We have mapped the transition region over a range of Mn fluxes from 0.5 to 8 x $10^{-9}$Torr for temperatures in the range 180 to 300$^{o}$C. The results of this study are shown in Figure 1. The apparent transition temperature varies slightly from one substrate holder to another and the most reproducible measure of growth temperature is obtained by using

RHEED to find the 3D/2D transition for any given Mn flux.

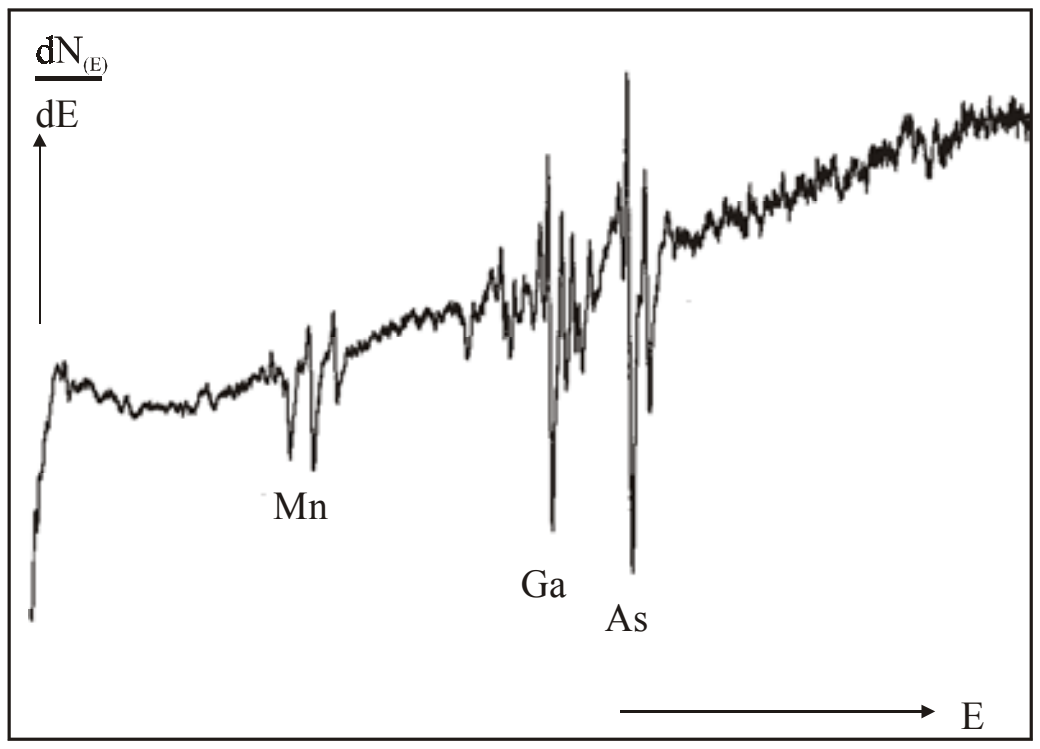

Figure 2 Auger electron spectroscopy trace for a GaMnAs film taken immediately after growth showing a high concentration of Mn on the surface.

Auger electron spectroscopy studies on films after growth show the presence of significant concentrations of Mn on the surface of the GaMnAs films. Figure 2 shows a typical example for a film grown in the 2D growth mode regime, the Mn peak heights are smaller than those of Ga and As, but comparable in intensity. Taking into account the information depth of the Auger signal, this suggests that the Mn surface concentration is in the range 0.1-1 monolayer, i.e. significantly higher than the bulk concentration. During growth in the 3D regime, the Mn concentration relative to Ga or As is even higher. Mn is thus behaving in a similar manner to all high vapour pressure dopants in GaAs [13] and to the higher vapour pressure species eg In in InGaAs [14]. It is well known that under such circumstances, there is an excess concentration of the higher vapour pressure species on the surface which increases until the incorporation into the bulk occurs at a rate that is equal to the arrival rate of that species. The ratio of surface to bulk concentration can be decreased slightly by increasing the As flux to the surface or more dramatically by lowering the growth temperature. This process is considered to be a kinetically hindered surface segregation mechanism. The behaviour of Mn is thus similar to many other species. With increasing Mn flux or increasing growth temperature, the Mn concentration on the surface during growth increases and at some critical value there is a transition from 2D to 3D growth mode associated with the high Mn surface concentration as shown in Figure 1.

### 3.2. Structural Quality

1μm thick GaMnAs films with nominal concentrations of 8, 5 and 2% Mn have been grown to study the structural quality of the material using X-ray diffraction. Figure 3 shows an asymmetric space map for 444 reflection from one of the films. This indicates that the structural quality of the GaMnAs film is comparable to that of the underlying substrate. More extensive studies of all the films, show that the mosaic block structure observed in some of the substrates is also present in the epitaxial films.

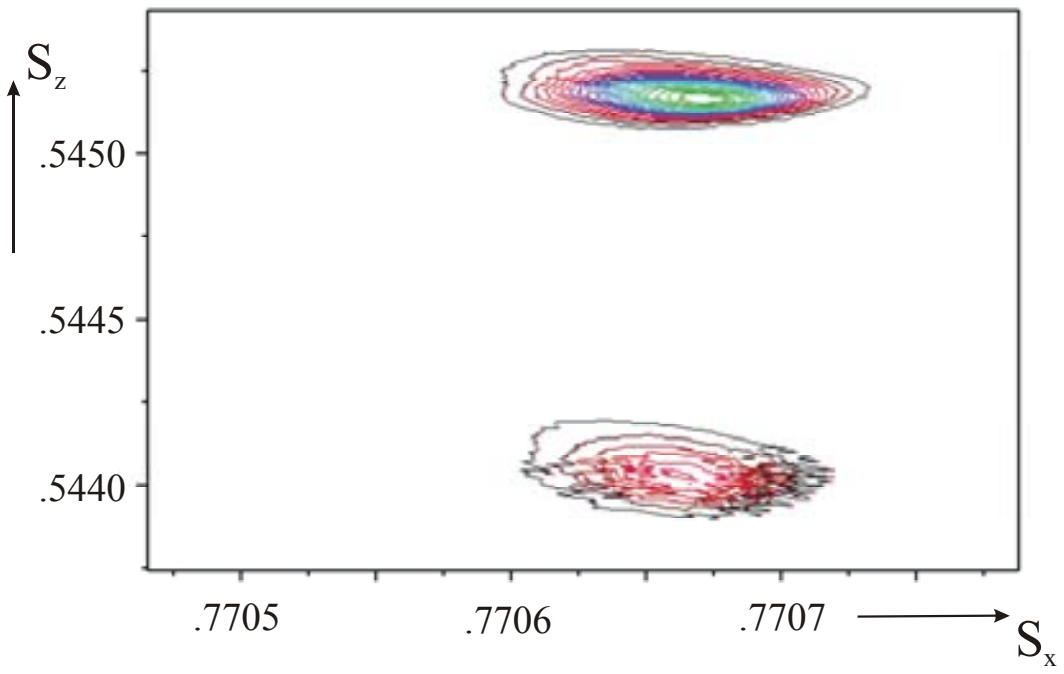

Figure 3 A reciprocal space map for a 1μm thick $Ga_{.98}Mn_{.02}As$ film taken using the 444 reflection. This demonstrates that the film is fully strained with respect to the GaAs substrate.

Our studies also show that all the GaMnAs films are fully strained with no detectable relaxation, thus indicating that no additional misfit dislocations are formed at the interface.

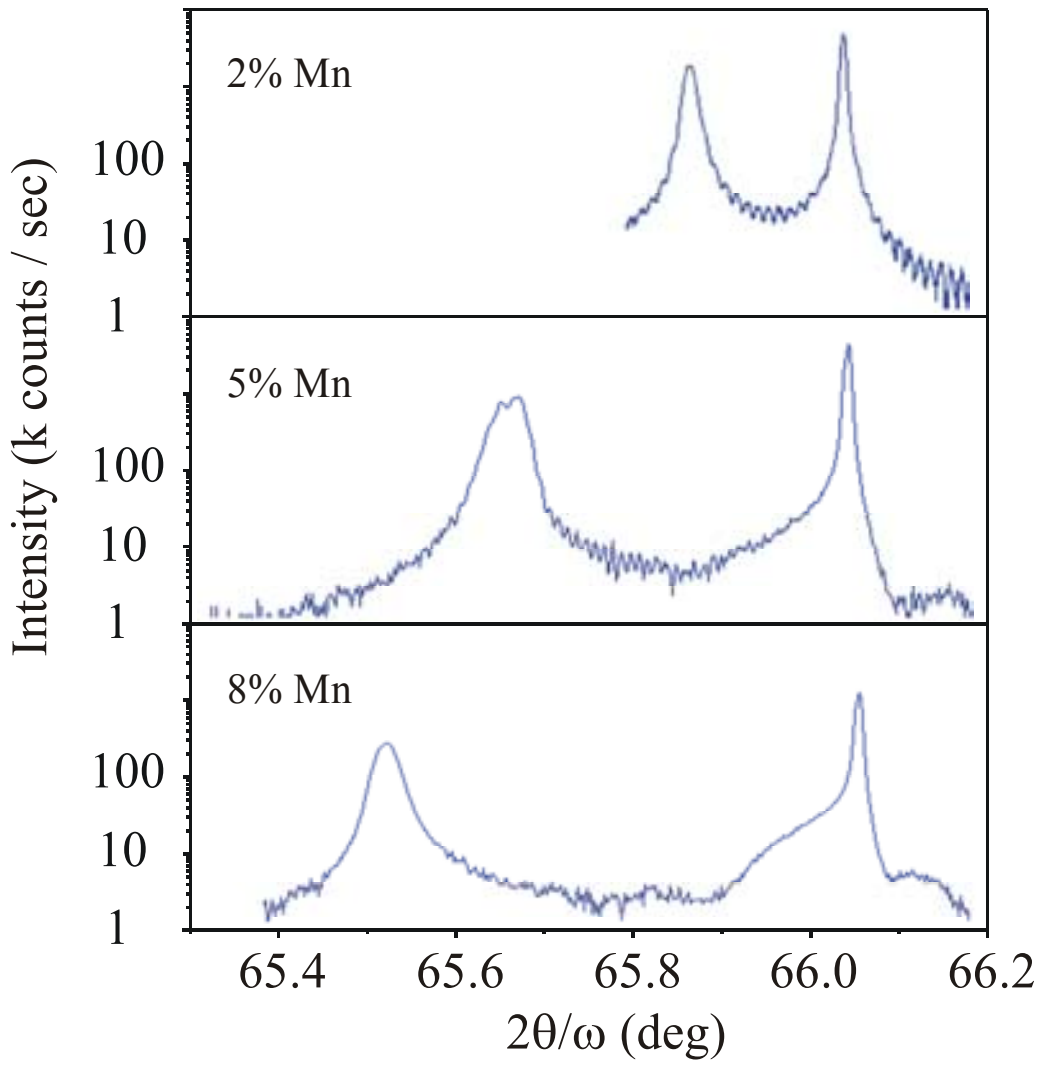

Figure 4 ω-2θ plots taken using 004 reflection for GaMnAs samples with nominal composition of 2% (a), 5% (b) and 8% (c) content.

Figure 4 shows ω-2θ plots for the 004 reflection for each of the 1μm thick GaMnAs samples. The lattice constant of the GaMnAs inceases systematically with increasing Mn content. By assuming Vegard's law is valid, the value of the lattice constant of a GaMnAs film has been used by many groups to deduce the Mn concentration. However this procedure has two major problems. Firstly the As antisite defects also lead to an increase in the lattice constant and secondly there is considerable uncertainty in the appropriate value of the lattice parameter of zinc-blende MnAs [15]. As we show below the density of As antisite defects is much lower for our $As_2$ grown samples than for samples grown with $As_4$, so this latter problem should be less severe. We have made independent estimates of the Mn concentration using XRF and EPMA measurements and have deduced the value for the lattice parameter of fully strained zinc-blende MnAs lattice matched to GaAs to be 0.613nm, this will be discussed elsewhere in detail [16]. Taking this value of lattice parameter, we obtain calculated compositions of 1.9, 5.3 and 8.3%, very close to the nominal values. Assuming Vegard's law is valid and a value of 0.311 for Poisson's ratio, the corresponding value for the relaxed lattice constant of zinc-blende MnAs is 0.591nm, which agrees closely with other values reported the literature [15]. The composition varies linearly with the Mn flux suggesting that Vegard's law is valid for this alloy family and that the value of the lattice parameter we have taken for MnAs is reasonable for such films. However independent estimates of the chemical concentration are required to test the validity of Vegard's law as is discussed in more detail elsewhere [16].

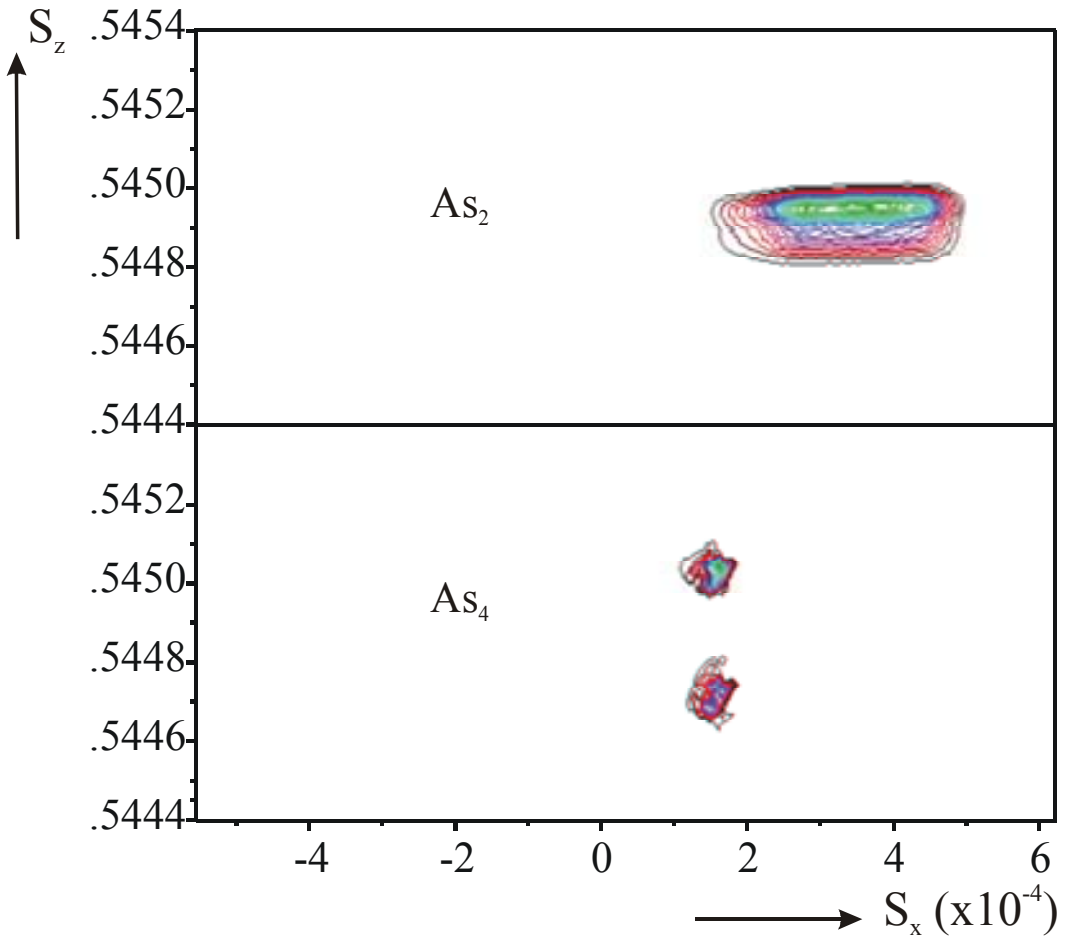

Figure 5 reciprocal space maps taken using a 004 reflection for films grown at 200°C using $As_2$ (a) and $As_4$ (b) respectively.

For comparison, 1μm thick LT GaAs films were grown at the same substrate temperatures used for the GaMnAs films. Figure 5 shows reciprocal space maps for two samples grown at the same temperature (200°C), one using $As_2$ and the other using $As_4$. The sample grown with the dimer shows only a single peak, ie the layer and substrate have the same lattice parameter, within experimental error. However, the corresponding film grown with the tetramer shows two distinct peaks indicating that the lattice parameter of the epitaxial layer is significantly different to that of the underlying substrate due to the higher concentration of antisite defects. This is in general observed in LT GaAs films grown with $As_4$ [17]. A double peak is observed in the $As_2$ film grown at the lowest temperature (175°C), but the results indicate that at any given growth temperature the density of antisite defects is lower for films grown with the dimer source of arsenic.

| LT GaAs Growth Temp °C | Substrate FWHM arc-sec | Layer FWHM arc-sec | As | No peaks |
|---|---|---|---|---|
| 175 | 12 | 22 | $As_2$ | 2 |
| 200 | 12 | 20 | $As_2$ | 1 |
| 200 | 12 | 18 | $As_4$ | 2 |
| 240 | 12 | 13 | $As_2$ | 1 |

| LT GaMnAs Growth Temp °C | Substrate FWHM arc-sec | Layer FWHM arc-sec | As | No peaks |
|---|---|---|---|---|
| 175 | 11 | 48 | $As_2$ | 2 |
| 200 | 13 | 67 | $As_2$ | 2 |
| 240 | 13 | 29 | $As_2$ | 2 |

Table 1 FWHM for the symmetric 004 reflection for the thick LT GaAs and GaMnAs samples

Table 1 shows the FWHM for the symmetric 004 reflection for the thick LT GaAs and GaMnAs samples. From this data it is evident that the FWHM for the GaMnAs films is larger than the value LT GaAs grown at the equivalent temperature. This strongly suggests that the defects introduced by LT growth using $As_2$ are small compared to those produced by the Mn. Comparing our data for the GaMnAs films with that present in the literature, the figures given in Table 1 are better than most of the values reported to date for equivalent Mn concentrations [18-20] and are comparable to the best reported values [21].

### 3.3 Electrical transport properties

The Hall resistance of ferromagnetic GaMnAs is dominated by the extraordinary Hall effect and thus may be used to obtain information on the magnetisation of the material. However it is usually difficult to extract the much smaller ordinary Hall effect and thus the carrier density. As will be discussed in detail elsewhere [22], we have be able to extract accurate values for the hole density in our sample by using temperature down to 300mK and magnetic fields up to 16.5Tesla. In part this results from the higher mobilities (1-5 $cm^2/Vs$) and lower resistivities (3-8 x $10^{-3}$ Ωcm) in our samples compared to those normally reported for films grown with $As_4$.

Our samples show metallic conductivity down to Mn concentrations as low as 1.5%. This contrasts with previous studies where authors report that GaMnAs is always insulating at such low concentrations. We find that of the degree of compensation is small for Mn concentration less than ~2% and that the hole concentration is close to the Mn concentration. This low compensation and resulting metallic conductivity is consistent with reduced levels of As antisite defects.

Figure 6 shows two typical plots for the saturation magnetisation versus temperature for a GaMnAs thin films.

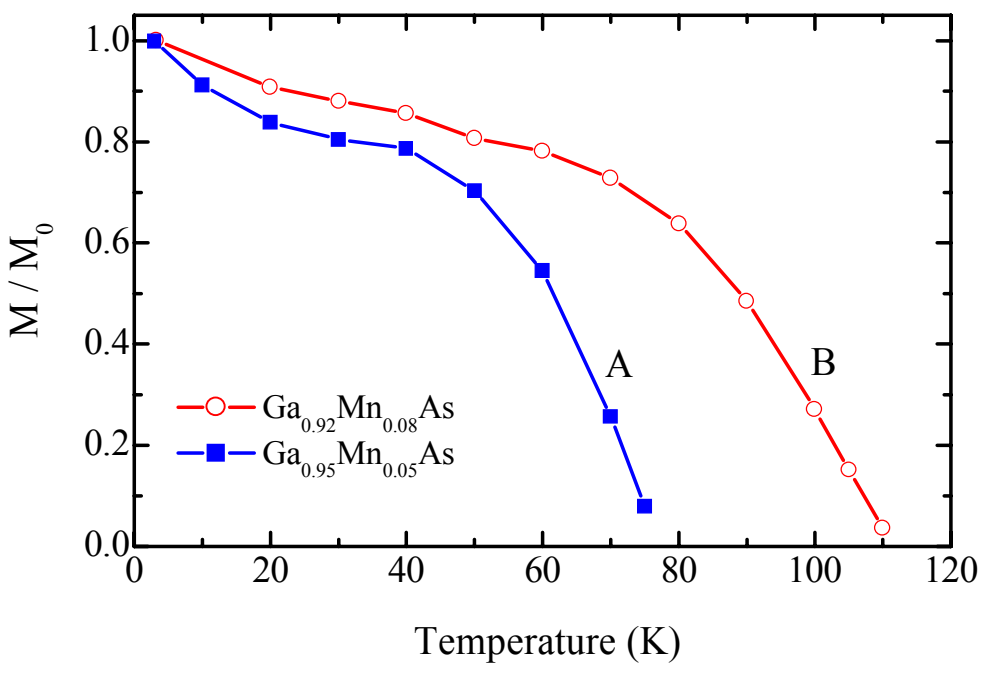

Figure 6 Saturation magnetisation curves versus temperature for an unannealed (A) and an annealed (B) sample of GaMnAs grown using $As_2$.

The Curie temperature, deduced from this data, is approximately 75K. Figure 6 also show a similar plot for a sample after in-situ annealing at the growth temperature. As is commonly observed, we see a considerable improvement in the Curie temperature to 112K as a result of annealing. This latter value is slightly higher than the best value reported to date for this system.

Our data is entirely consistent with that reported recently by Sadowski et al [23], who also used arsenic dimer source, but reported on the structural quality of their films. In that study Mn concentrations of up to 3.5% resulted in films with a Curie temperature of 50K somewhat lower than the values we observe for similar Mn concentrations.

## 4. Conclusions

In conclusion, GaMnAs films grown with $As_2$ show excellent structural, electrical and magnetic properties, which are at least in part due to a lower concentration of antisite defects in thin films grown with the dimer. Films showing metallic conduction for a wide range of Mn concentrations from 1.5 to 8% are obtained. Curie temperatures comparable or better than those previously reported are readily obtained using $As_2$.

## Acknowledgements

The authors would like to acknowledge the support for this work from EPSRC (Grant Number GR/ R17652/01) and the EU (FENIKS project EC: G5RD-CT-2001-00535). The authors would also like to thank the L Eaves, M Henini, P C Main and A Patane for helpful discussion and J Chauhan, B Hill, J Middleton and A Neumann for technical support.